\documentclass[12pt]{article}

\usepackage{epsfig}
\usepackage{amsmath}
\usepackage{amssymb}
\usepackage{axodraw}

\providecommand{\ap}{{\alpha^\prime}}

\providecommand{\ap}{{\alpha^\prime}}

\providecommand{\Dta}{{\overline{D3}}}

\begin{document}

\begin{titlepage}

\begin{flushright}
UMD-PP-03-mmm\\
%\today
\end{flushright}

\vspace{1cm}

\begin{center}
\baselineskip25pt {\Large\bf Dual Branes, Discrete Chain States and the
Entropy of the Schwarzschild Black Hole}
\end{center}

\vspace{1cm}

\begin{center}
{Axel Krause\footnote{E-mail: {\tt
krause@physics.umd.edu}}}\\[10mm]
{\it Center for String and Particle Theory}\\[1.8mm]
{\it Department of Physics, University of Maryland}\\[1.8mm]
{\it College Park, MD 20742-4111, USA}
\vspace{0.3cm}
\end{center}

\vspace*{\fill}

\begin{abstract}
We review a recent proposal towards a microscopic understanding
of the entropy of non-supersymmetric spacetimes -- with emphasis
on the Schwarzschild black hole. The approach is based at an
intermediate step on the description of the non-supersymmetric
spacetime in terms of dual Euclidean brane pairs of type-II
string-theory or M-theory. By counting specific chain structures
on the brane-complex, it is shown that one can reproduce the
exact Schwarzschild black hole entropy plus its logarithmic
correction.
\end{abstract}
\vspace*{\fill}
\end{titlepage}

\section{Introduction}
We know since the early work of
\cite{Chr},\cite{Haw1},\cite{TDBH},\cite{Bek},\cite{Haw2} that
black holes come equipped with an entropy, known as the
Bekenstein-Hawking (BH) entropy. It is determined by the area
$A_H$ of the hole's event horizon
\begin{equation}
{\cal S}_{BH} = \frac{A_H}{4G_4}\frac{k_Bc^3}{\hbar} \; .
\end{equation}
Since the laws of black hole thermodynamics are in a neat
one-to-one correspondence with the conventional laws of
thermodynamics, one expects in analogy that the BH-entropy should
likewise be explainable statistical mechanically by counting the
entropy of an underlying set of microstates in a microcanonical
ensemble.

So far string-theory satisfied these expectations for a variety
of supersymmetric black holes (see e.g.~\cite{REV} for reviews).
Here, supersymmetry is an essential ingredient given that the
counting of the microscopic degrees of freedom takes place in a
different regime of moduli space than where the black hole
actually resides \cite{SV}. One counts the entropy of specific
string excitations at weak string coupling where the string
dynamics is under control and by relying on supersymmetry exports
the result to the strongly coupled regime where the black hole
lives but where up to now the full string-theory dynamics is not
sufficiently understood. In another approach which tries to avoid
the supersymmetry constraint one maps the spacetime of interest
by means of T- and S-Dualities and sometimes also boost
transformations to a spacetime whose microscopic entropy counting
is under control, e.g.~the D=3 BTZ black hole or near-extremal
branes \cite{SS}. Due to lack of our knowledge of the full
formulation of string-/M-theory in the strongly coupled regime,
both methods rely on an indirect method of counting the relevant
degrees of freedom. It is therefore not obvious what the degrees
of freedom actually are that make up the black hole in the
strongly coupled regime. In particular it is not clear where
these are located.

Our aim here is to introduce an alternative approach which does
not rely on supersymmetry and therefore might be applied to both
non-supersymmetric and supersymmetric spacetimes. Moreover we do
not exploit weakly coupled string technology for the entropy
counting in order not to be constrained from the outset to this
region in moduli space. Instead the intention of this approach is
to focus directly on the entropy counting problem in the strongly
coupled regime where the string-coupling constant $g_s$ becomes
${\cal O}(1)$. Of course since we still lack a complete
formulation of string-/M-theory in this regime, the idea is to
solve the state-counting issue first, thereby extracting a
suitable set of microstates for the strongly coupled regime and
then in a future second step to build the dynamics on these
states in order to hopefully arrive at a non-perturbative
formulation of string-theory. The presentation given here
focusses on a proposition and counting of states for a D=4
spacetime with spherical event horizon and is based on
\cite{K11},\cite{K12},\cite{K13}. Our strategy will be to
reformulate the D=4 BH-entropy purely in terms of geometrical
Nambu-Goto actions corresponding to certain dual brane pairs and
then to consider suitable microstates related to these branes
whose entropy will be compared to the BH-entropy.

\section{BH-Entropy and Dual Brane Doublets}
Let us consider type II String-Theory on a D=10 Lorentzian
spacetime ${\cal M}^{1,3}\times {\cal M}^{p-1}\times {\cal
M}^{7-p}$ ($p=1,\hdots,5$) with D=4 external part ${\cal M}^{1,3}$
\begin{equation}
ds^2 = g_{\mu\nu}^{(1,3)}dx^\mu dx^\nu +
g_{ab}^{(p-1)}(x^c)dx^adx^b + g_{kl}^{(7-p)}(x^m)dx^kdx^l \; .
\end{equation}
We consider ${\cal M}^{p-1}$ and ${\cal M}^{7-p}$ as compact such
that we have a geometric background describing a compactification
from D=10 down to D=4. For such a background the D=4 Newton's
constant is related to the Regge slope $\ap$ and $g_s$ through
\begin{equation}
G_4=\frac{G_{10}}{V_{p-1}V_{7-p}}=\frac{(2\pi)^6\ap^4g_s^2}{8V_{p-1}
V_{7-p}} \; ,
\label{NEWTON}
\end{equation}
where $V_i=vol({\cal M}^{i})\equiv\int_{{\cal M}^{i}}d^ix
\sqrt{g^{(i)}}$.

Next, let us choose a sphere $S^2$ within the spacelike part of
${\cal M}^{1,3}$ and wrap two mutually orthogonal Euclidean
``electric-magnetic'' dual branes, $Dp$ and $D(6-p)$, around
$S^2\times{\cal M}^{p-1}$ and ${\cal M}^{7-p}$, resp.~such that
together they cover the whole internal space plus the external
sphere. For such a dual brane pair it follows from the
Dirac-quantization condition that the product of their tensions
obeys
\begin{equation}
\tau_{Dp}\tau_{D(6-p)}=\frac{1}{(2\pi)^6\ap^4g_s^2} \; .
\end{equation}
Thus we can write the inverse of the D=4 Newton's constant as
\begin{equation}
\frac{1}{G_4}=8(\tau_{Dp}V_{p-1})(\tau_{D(6-p)}V_{7-p}) \; .
\end{equation}
Notice that the right-hand-side of this expression comes already
close to the product of two Nambu-Goto actions for the
resp.~Euclidean branes except for the fact that we miss the area
of the external two-sphere in the first bracket.

Since our aim is here to deal with the BH-entropy of D=4
spacetimes with spherical event horizons $S^2_H$, let us now point
out how we incorporate them and what is the role played by the
dual branes. In general the branes will act as supergravity
sources and therefore will give rise to some particular D=10
geometry of the form
\begin{equation}
(\text{D=4 Spacetime})\times(\text{Compact Internal Space}) \; .
\end{equation}
If we include no other sources than the dual branes, the D=4
geometry will possess spherical symmetry because the D=4
gravitational source is distributed evenly over the $S^2$. So we
have to ask whether there exists suitable dual branes (actually
we will see soon that what we need is a doublet of dual brane
pairs) whose D=10 geometry includes the D=4 spacetime of interest
in its external part. For what follows, it will be important
moreover to have an identification of the $S^2$ with the event
horizon sphere $S^2_H$ of the D=4 spacetime
\begin{equation}
S^2\equiv S^2_H  \; . \label{equiv}
\end{equation}
Under this hypothesis for which evidence for the Schwarzschild
black hole case was given in \cite{K12},\cite{K13}, we will now
proceed with the general argument and will comment below more on
the specific brane pairs needed for the D=4 Schwarzschild black
hole description.

Assuming the identification (\ref{equiv}) it is easy to see that
by using (\ref{NEWTON}) we can reformulate the D=4 spacetime's
BH-entropy as
\begin{equation}
{\cal S}_{BH}=\frac{A_H}{4G_4}=2S_{Dp}S_{D(6-p)} \; ,
\end{equation}
where
\begin{equation}
S_{Dp}=\tau_{Dp}\int_{S^2\times{\cal M}^{p-1}} d^{p+1}x\sqrt{\det
g} \; , \qquad
S_{D(6-p)}=\tau_{D(6-p)}\int_{{\cal M}^{7-p}}
d^{7-p}x\sqrt{\det g}
\end{equation}
are the respective Nambu-Goto actions of the involved dual branes.
Actually, we can also get rid of the factor two by repeating the
procedure once more and employing a doublet of Euclidean brane
pairs. This doubling of the dual pairs will turn out to be
important in the second half when we determine the microscopic
entropy. Notice, that there is no {\em a priori} reason why the
second dual pair has to coincide with the first. Therefore let us
wrap a further dual Euclidean brane pair $Dp'-D(6-p')$, with
$Dp'$ and $D(6-p')$ again mutually orthogonal, in the same manner
as before around the $S^2$ plus the internal space. Then by
following the previous steps, the D=4 BH-entropy becomes purely
expressible in terms of the respective Nambu-Goto actions
\begin{equation}
{\cal S}_{BH}=S_{Dp}S_{D(6-p)}+S_{Dp'}S_{D(6-p')} \; .
\end{equation}
Furthermore, notice that we are free to exchange any of the
appearing branes with its anti-brane and still arrive at the same
expression (of course under the premise that the inclusion of the
antibrane gives rise to the requested D=4 spacetime in the
external part of the ensuing D=10 geometry).

It turns out that this formula works for all doublets of dual
brane pairs of string- and M-Theory \cite{K11}. Thus whenever we
find a doublet of dual brane pairs $(E_1,M_1),(E_2,M_2)$ which
gives rise to the requested D=4 spacetime with spherical event
horizon in the external part of the D=10 metric together with the
identification (\ref{equiv}), it is possible to rewrite the
BH-entropy of the D=4 spacetime as
\begin{equation}
{\cal S}_{BH}=\sum_{i=1,2}S_{E_i}S_{M_i} \; , \label{BHEnt}
\end{equation}
where the pairs $(E_i,M_i)$ range over all possible dual pairs of
type-II string-theory and M-theory
\begin{equation}
(E_i,M_i) \in \{(Dp_i,D(6-p_i)),(F1,NS5),(NS5,F1),(M2,M5),
(M5,M2)\} \; .
\label{ADO}
\end{equation}
Moreover, any occuring object might also be replaced by its
anti-partner as this exchange leaves the Nambu-Goto action
invariant.

In particular, for the case of the D=4 Schwarzschild black hole
one might take the self-dual $(D3,D3),(\Dta,\Dta)$ type-IIB brane
anti-brane doublets, distributed along the ten coordinates as
depicted in fig.\ref{setup}.
\setcounter{figure}{0}
\begin{figure}[t]
\begin{center}
\begin{picture}(230,85)(0,-10)
\Text(30,72)[]{$t$}
\Text(50,72)[]{$r$}
\Text(70,72)[]{$\theta$}
\Text(90,72)[]{$\phi$}
\Text(110,72)[]{$4$}
\Text(130,72)[]{$5$}
\Text(150,72)[]{$6$}
\Text(170,72)[]{$7$}
\Text(190,72)[]{$8$}
\Text(210,72)[]{$9$}
\Text(2,54)[]{$D3$}
\Text(2,36)[]{$\Dta$}
\Text(2,18)[]{$D3$}
\Text(2,0)[]{$\Dta$}
\Line(-3,63)(215,63)
\Line(17,-6)(17,74)
\Text(70,54)[]{$\bullet$}
\Text(90,54)[]{$\bullet$}
\Text(110,54)[]{$\bullet$}
\Text(130,54)[]{$\bullet$}
\Text(70,36)[]{$\bullet$}
\Text(90,36)[]{$\bullet$}
\Text(110,36)[]{$\bullet$}
\Text(130,36)[]{$\bullet$}
\Text(150,18)[]{$\bullet$}
\Text(170,18)[]{$\bullet$}
\Text(190,18)[]{$\bullet$}
\Text(210,18)[]{$\bullet$}
\Text(150,0)[]{$\bullet$}
\Text(170,0)[]{$\bullet$}
\Text(190,0)[]{$\bullet$}
\Text(210,0)[]{$\bullet$}
\end{picture}
\caption{The Euclidean brane anti-brane pairs
$(D3,D3),(\Dta,\Dta)$ which describe a D=4 Schwarzschild black
hole. They are oriented along the directions marked by dots. The
coordinates $t,r,\theta,\phi$ describe the external D=4 spacetime
with $\theta,\phi$ describing the sphere $S^2=S^2_H$. The whole
configuration is located at some common fixed D=4 radial value
$r=r_H$ describing the event horizon of the black hole in
Schwarzschild coordinates.}
\label{setup}
\end{center}
\end{figure}
An equal amount of branes and anti-branes guarantees an uncharged
solution of the D=10 supergravity field equations and moreover
leads to a non-supersymmetric background. Furthermore, the choice
of the non-dilatonic $D3$'s ensures that one finds a D=10 vacuum
solution without dilaton matter present. It can then be worked out
\cite{K12} that indeed the described $(D3,D3),(\Dta,\Dta)$
configuration gives rise in the D=4 part of the metric to an
exterior Schwarzschild black hole geometry outside the $S^2$
together with the identification of the spheres (\ref{equiv}).
Notice that Euclidean branes positioned somewhere in spacetime
usually decay because they are localized in time. However,
Euclidean branes which are positioned at an event horizon are
seen by an outside (for whom $r$ is bigger than $r_H$ -- the
position of the brane configuration) D=4 observer through an
infinite redshift. This stretches classically any decay-time
infinitely long such that from the observer's perspective
Euclidean branes wrapping an event horizon lead to a stationary
spacetime \cite{K12}.

In this context it might be interesting to note as an aside that
brane-antibrane systems in general cannot annihilate and decay
into the closed-string vacuum classically \cite{SenRad}. Such a
decay in which the tachyon would roll down its potential hill and
thereby radiating off into the bulk the surplus of energy can
occur only quantum mechanically. In weakly coupled string-theory
this can be understood from the fact that during this transition
open strings on the brane-antibrane system have to transform into
closed strings which can enter the bulk. However this is a
one-loop process from the open string point of view and thus a
quantum mechanical process. On the other hand it is well-known
that also the black hole's Hawking radiation is an intrinsic
quantum mechanical phenomenon which is forbidden at the classical
level and likewise describes radiation sent into the bulk. It
therefore seems natural to conjecture within the D=10
$(D3,D3),(\Dta,\Dta)$ description of the D=4 Schwarzschild black
hole that its {\em Hawking radiation might be due to tachyon
condensation} \cite{K12}. It would certainly be interesting to
investigate this further which however we will not do here.

\section{Chain-States and their Entropy}
To proceed further with the analysis of the D=4 BH-entropy, let
us reflect briefly upon the tension of a Euclidean brane. For
Lorentzian $Dp$-branes their tension is usually interpreted as
\begin{equation}
\tau_{Dp}=\frac{\text{mass}}{\text{unit of p-volume}}
\end{equation}
which treats the time and the space directions differently. For a
Euclidean brane however there is no time direction on the
worldvolume and instead one has to treat all spacelike worldvolume
directions on an equal footing. It therefore seems required in
this case to interpret the tension
\begin{equation}
\tau_{Dp}=:\frac{1}{v_{Dp}}= \frac{1}{l_{Dp}^{p+1}}
\label{WVU}
\end{equation}
in terms of a smallest fundamental volume unit $v_{Dp}$ giving a
smallest length $l_{Dp}$. This is a natural generalization of the
fact that $\sqrt{\ap}$ constitutes a smallest length for strings.
In the strongly coupled regime, however, it will be chains -- to
be introduced shortly -- which cannot resolve distances smaller
than $l_{Dp}$.

Evidence for such a smallest volume unit on the brane's
worldvolume comes from the ``worldvolume uncertainty relations
for D-branes'' \cite{CHK}. In \cite{CHK} it was shown that the
worldvolume $X^0,\hdots,X^p$ of a $Dp$-brane (which could be
Lorentzian or Euclidean) is subject to the following uncertainty
relation
\begin{equation}
Dp\; : \quad \delta X^0 \delta X^1 \hdots \delta X^p \gtrsim g_s
{\ap}^{\frac{p+1}{2}} \; ,
\end{equation}
where the right-hand-side was determined up to numerical factors.
By employing string-dualities it was furthermore shown that
similar uncertainty relations hold true for the $NS5$-brane, the
fundamental string and the M-theory $M5$- and $M2$-branes
\begin{alignat}{3}
NS5\; &: \quad \delta X^0 \delta X^1 \hdots \delta X^5 \gtrsim
g_s^2 {\ap}^3 \; , \qquad\qquad
F1\; &&: \quad \delta X^0 \delta X^1 \gtrsim \ap \; , \\
M5\; &: \quad \delta X^0 \delta X^1 \hdots \delta X^5 \gtrsim
l_{Pl}^6 \; , \qquad\qquad\quad M2\; &&: \quad \delta X^0 \delta
X^1 \delta X^2 \gtrsim l_{Pl}^3 \; ,
\end{alignat}
with $l_{Pl}$ the D=11 Planck-length. Indeed the relation for the
fundamental string had been proposed earlier in \cite{Yo}. By
employing the tensions of these objects
\begin{alignat}{3}
Dp &: \; \tau_{Dp} = \frac{1}{(2\pi)^p g_s \ap^{(p+1)/2}} \; ,
\quad
NS5 &&: \; \tau_{NS5} = \frac{1}{(2\pi)^5 g_s^2 \ap^3}\; ,
\quad
F1 : \; \tau_{F1} = \frac{1}{2\pi \ap} \; , \\
M5 &: \; \tau_{M5} = \frac{1}{(2\pi)^5 l_{Pl}^6} \; , \qquad\qquad
M2 &&: \; \tau_{M2} = \frac{1}{(2\pi)^2 l_{Pl}^3}
\end{alignat}
one sees that all the different worldvolume uncertainty relations
can be combined into the statement that the smallest worldvolume
allowed by the brane uncertainty principle is given by the inverse
of the object's tension
\begin{equation}
\delta X^0 \hdots \delta X^p \gtrsim \frac{1}{\tau} \; .
\end{equation}
This motivates us to interpret similarly the tension $\tau_E$ or
$\tau_M$ of any of the dual objects occuring in (\ref{ADO})
analogously to (\ref{WVU}) in terms of some smallest volume
$v_{E,M}=\tau_{E,M}^{-1}$.

The introduction of a smallest volume unit on the brane's
worldvolume endows the brane with a discrete structure. Namely,
we can now perceive the brane as a lattice made out of a certain
number $N_{Dp}$ of such smallest volume units which we will call
cells from on. With the interpretation (\ref{WVU}) for the
tension, it is precisely the Nambu-Goto action which measures the
number of these cells contained in the brane
\begin{equation}
N_{Dp}:=\tau_{Dp}\int d^{p+1}x\sqrt{\det g}=S_{Dp} \; .
\end{equation}
This implies that the expression (\ref{BHEnt}) which we had found
for the D=4 BH-entropy can be rewritten further and now becomes
identical to an integer $N$
\begin{equation}
{\cal S}_{BH}=\sum_{i=1,2}N_{E_i}N_{M_i} =: N
\label{BHEnt2}
\end{equation}
with $N$ the total number of cells contained in the joint
worldvolume of the doublet of dual brane pairs.

So far we have reformulated the D=4 BH-entropy in terms of
string-/M-theory notions under the inclusion of the discrete brane
structure coming from a reinterpretation of the brane's tension.
We will now investigate in which way this helps us to propose a
set of black hole microstates for the strongly-coupled regime.

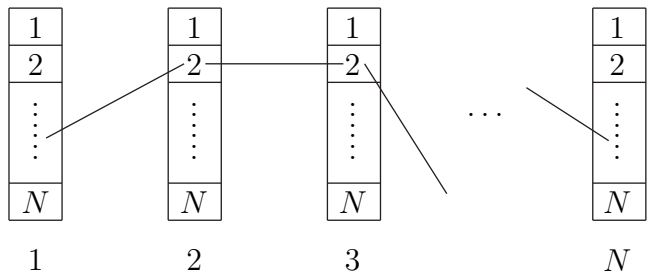
\begin{figure}[t]
\begin{center}
\begin{picture}(260,120)(0,0)
\Text(10,5)[]{$1$}
\Line(0,20)(0,100)
\Line(20,20)(20,100)
\Line(0,100)(20,100)
\Line(0,86)(20,86)
\Line(0,72)(20,72)
\Line(0,34)(20,34)
\Line(0,20)(20,20)
\Text(10,93)[]{$1$}
\Text(10,79)[]{$2$}
\Text(10,63)[]{$\vdots$}
\Text(10,51)[]{$\vdots$}
\Text(10,27)[]{$N$}
\Line(14,51)(66,79)
\Text(70,5)[]{$2$}
\Line(60,20)(60,100)
\Line(80,20)(80,100)
\Line(60,100)(80,100)
\Line(60,86)(80,86)
\Line(60,72)(80,72)
\Line(60,34)(80,34)
\Line(60,20)(80,20)
\Text(70,93)[]{$1$}
\Text(70,79)[]{$2$}
\Text(70,63)[]{$\vdots$}
\Text(70,51)[]{$\vdots$}
\Text(70,27)[]{$N$}
\Line(74,79)(126,79)
\Text(130,5)[]{$3$}
\Line(120,20)(120,100)
\Line(140,20)(140,100)
\Line(120,100)(140,100)
\Line(120,86)(140,86)
\Line(120,72)(140,72)
\Line(120,34)(140,34)
\Line(120,20)(140,20)
\Text(130,93)[]{$1$}
\Text(130,79)[]{$2$}
\Text(130,63)[]{$\vdots$}
\Text(130,51)[]{$\vdots$}
\Text(130,27)[]{$N$}
\Line(134,79)(165,30)
\Text(180,60)[]{$\hdots$}
\Text(230,5)[]{$N$}
\Line(220,20)(220,100)
\Line(240,20)(240,100)
\Line(220,100)(240,100)
\Line(220,86)(240,86)
\Line(220,72)(240,72)
\Line(220,34)(240,34)
\Line(220,20)(240,20)
\Text(230,93)[]{$1$}
\Text(230,79)[]{$2$}
\Text(230,63)[]{$\vdots$}
\Text(230,51)[]{$\vdots$}
\Text(230,27)[]{$N$}
\Line(195,70)(226,50)
\end{picture}
\caption{Constructive view of an $(N-1)$-chain where we arrange
all cells of the lattice in a column and use $N$ copies of them.
We allow each link to connect any cell of a column with any cell
of the succeeding column. Horizontal links correspond to loops.}
\label{openchains}
\end{center}
\end{figure}

\begin{figure}[t]
\begin{center}
\begin{picture}(260,160)(0,10)
\Line(30,35)(30,115)
\Line(50,35)(50,115)
\Text(40,127)[]{$2$}
\Line(30,115)(50,115)
\Text(40,108)[]{$1$}
\Line(30,101)(50,101)
\Text(40,94)[]{$2$}
\Line(30,87)(50,87)
\Text(40,78)[]{$\vdots$}
\Text(40,66)[]{$\vdots$}
\Line(30,49)(50,49)
\Text(40,42)[]{$N$}
\Line(30,35)(50,35)
\Line(60,70)(60,150)
\Line(80,70)(80,150)
\Text(70,162)[]{$3$}
\Line(60,150)(80,150)
\Text(70,143)[]{$1$}
\Line(60,136)(80,136)
\Text(70,129)[]{$2$}
\Line(60,122)(80,122)
\Text(70,113)[]{$\vdots$}
\Text(70,101)[]{$\vdots$}
\Line(60,84)(80,84)
\Text(70,77)[]{$N$}
\Line(60,70)(80,70)
\Line(88,10)(88,90)
\Line(108,10)(108,90)
\Text(98,102)[]{$1$}
\Line(88,90)(108,90)
\Text(98,83)[]{$1$}
\Line(88,76)(108,76)
\Text(98,69)[]{$2$}
\Line(88,62)(108,62)
\Text(98,53)[]{$\vdots$}
\Text(98,41)[]{$\vdots$}
\Line(88,24)(108,24)
\Text(98,17)[]{$N$}
\Line(88,10)(108,10)
%% Die Verbindungslinien %%
\Line(120,120)(73,129)
\Line(67,129)(40,60)
\Line(40,60)(98,52)
\Line(98,52)(162,28)
\Line(162,28)(220,63)
\Line(220,63)(190,110)
\Line(190,110)(140,135)
\Text(130,130)[]{$\hdots$}
%%%%%%%%%%%%%%%%%%%%%%%%%%%
\Line(152,10)(152,90)
\Line(172,10)(172,90)
\Text(162,102)[]{$N$}
\Line(152,90)(172,90)
\Text(162,83)[]{$1$}
\Line(152,76)(172,76)
\Text(162,69)[]{$2$}
\Line(152,62)(172,62)
\Text(162,53)[]{$\vdots$}
\Text(162,41)[]{$\vdots$}
\Line(152,24)(172,24)
\Text(162,17)[]{$N$}
\Line(152,10)(172,10)
\Line(180,70)(180,150)
\Line(200,70)(200,150)
\Text(190,162)[]{$N-2$}
\Line(180,150)(200,150)
\Text(190,143)[]{$1$}
\Line(180,136)(200,136)
\Text(190,129)[]{$2$}
\Line(180,122)(200,122)
\Text(190,113)[]{$\vdots$}
\Text(190,101)[]{$\vdots$}
\Line(180,84)(200,84)
\Text(190,77)[]{$N$}
\Line(180,70)(200,70)
\Line(210,35)(210,115)
\Line(230,35)(230,115)
\Text(220,127)[]{$N-1$}
\Line(210,115)(230,115)
\Text(220,109)[]{$1$}
\Line(210,102)(230,102)
\Text(220,95)[]{$2$}
\Line(210,88)(230,88)
\Text(220,79)[]{$\vdots$}
\Text(220,67)[]{$\vdots$}
\Line(210,49)(230,49)
\Text(220,42)[]{$N$}
\Line(210,35)(230,35)
\end{picture}
\caption{Alternatively one might use closed $N$-chains. They possess the same
number of configurations.}
\label{closedchains}
\end{center}
\end{figure}
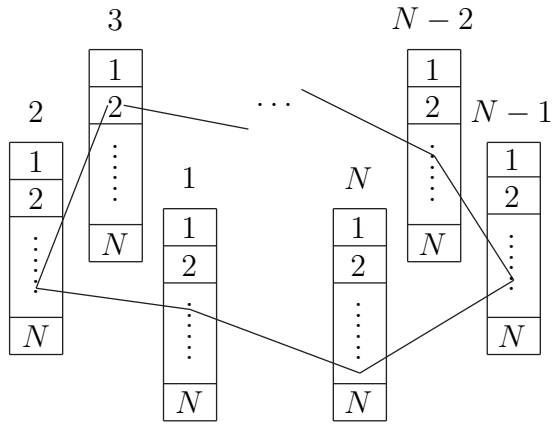

To this aim, let us conceive on the combined $(E_1,M_1),(E_2,M_2)$
worldvolume lattice an $(N-1)$-chain, which is a chain composed
out of $N-1$ successive links where we allow all links to start
and end democratically on any of the $N$ cells of the lattice
(see fig.\ref{openchains}).
In particular a link might start and end on the same cell thus
creating a loop. Altogether the number of possible chain
configurations is $N^N$. Alternatively, one might consider closed
$N$-chains (see fig.\ref{closedchains})
which exhibit the same number $N^N$ of different chain
configurations.

Part of the motivation to consider long chains (long enough to
connect all cells of the whole lattice instead of just a few)
comes from the weakly coupled string at finite temperature. So
far, we had argued for a smallest unresolvable volume on the
Euclidean branes -- or in other words an uncertainty in
space-resolution -- which leads via the Heisenberg uncertainty
principle to an uncertainty in energy (where we assume $l_{Dp}$
small enough so that $\Delta P$ becomes relativistic)
\begin{equation}
\Delta E \simeq \Delta P \simeq
\frac{1}{l_{Dp}}=(\tau_{Dp})^{\frac{1}{p+1}}
=\frac{1}{\sqrt{\ap}(g_s(2\pi)^p)^{\frac{1}{p+1}}}  \; .
\end{equation}
Therefore at strong coupling where $g_s\simeq 1$, the temperature
associated with the uncertainty $\Delta E$ is of order the
Hagedorn-temperature. Now at this temperature we know that in the
weakly coupled regime it is entropically favourable to allocate
the energy of the system to just one single long string instead to
many short ones \cite{AA}. This motivates us to consider as well
long chains instead of short ones for the strongly coupled
regime. Moreover it is known that open and closed strings behave
at high excitation levels much like a random walk
\cite{MT} which furthermore motivated the choice of
discrete chain states.

The counting of different chain configurations has up to now been
classical in the sense that all cells were regarded as
distinguishable. However, in a proper quantum theory the cells
should likely be regarded as indistinguishable bosonic degrees of
freedom. How to account for this quantum feature is well-known
from statistical mechanics -- for $N$ indistinguishable cells we
have to divide by the Gibbs-correction factor $N!$. This gives
the quantum-mechanically corrected number of different chain
states
\begin{equation}
\Omega (N) =\frac{N^N}{N!} \; .
\end{equation}
To evaluate the entropy of the chain-states in the thermodynamic
large $N$ limit (as adequate for macroscopic black holes) we use
Stirling's approximation, $\ln(N!)=N\ln N-N +{\cal O}(\ln N)$,
and the identity for the D=4 BH-entropy (\ref{BHEnt2}) to obtain
\begin{equation}
{\cal S}_{c} = \ln\Omega(N) = N = {\cal S}_{BH}
\end{equation}
up to corrections of ${\cal O}(\ln N)$. Thus we learn that the
proposed discrete $(N-1)$-chains or alternatively the closed
$N$-chains correctly reproduce the corresponding D=4 BH-entropy
and therefore might be considered as viable black hole microstate
candidates.

\section{Corrections to BH-Entropy}
Having found agreement between the chain and the D=4 BH-entropy
at leading large $N$ order, one might be curious what happens at
subleading order. Corrections to the BH-entropy for D=4 black
holes had been determined in supersymmetric cases from
String-Theory while results in non-supersymmetric cases came from
the Quantum Geometry program \cite{KM} or the conformal field
theory approach of Carlip \cite{CLC}.

The general result is that there exists a subleading logarithmic
correction to the semiclassical BH-entropy of the form
\begin{equation}
-k\ln{\cal S}_{BH}
\end{equation}
with a positive constant $k>0$ (a negative subleading correction
is in accordance with the holographic principle). Though
initially a value of $k=3/2$ had been favoured, it seems that
recently this has been corrected to $k=1/2$ \cite{JY}. The reason
being that one obtains a correction of the form
\begin{equation}
-\frac{3}{2}\ln{\cal S}_{BH}+\ln c \; .
\end{equation}
However, the central charge $c$ has been shown in \cite{JY} not
to be constant but instead given by $c \propto {\cal S}_{BH}$
which leads to an entropy correction with $k=1/2$. Moreover,
$k=1/2$ has also been found independently by other methods
exploiting the AdS/CFT duality \cite{MP}.

Let us now come to the entropy corrections within the chain state
approach. Corrections to the chain entropy come simply from a more
accurate approximation of $N!$ by the Stirling-series,
e.g.~including higher-order terms one could take
\begin{equation}
N!=\sqrt{2\pi N}N^Ne^{-N}\big( 1+\frac{1}{12N}
+{\cal O}\big(\frac{1}{N^2}\big) \big) \; .
\end{equation}
Using this corrected Stirling approximation for the evaluation of
the chain entropy and once more considering the identity
(\ref{BHEnt2}), we get a corrected chain-entropy formula
\begin{equation}
{\cal S}_c=\ln\Omega(N)
 ={\cal S}_{BH}-\frac{1}{2}\ln{\cal S}_{BH}
-\ln\sqrt{2\pi}-\frac{1}{12{\cal S}_{BH}} +{\cal O}\big(
\frac{1}{{\cal S}_{BH}^2} \big) \; .
\end{equation}
This shows that the proposed chain configurations can also easily
account for the subleading logarithmic entropy correction and
moreover give the precise numerical coefficient $k=1/2$. We can
therefore conclude that the proposed chain states have passed a
first non-trivial test and consequently constitute an interesting
possible set of black hole microstates whose dynamics should be
worth investigating in more detail.

We would like to stress that the mechanism of counting the chain
entropy is not restricted to some specific value of $g_s$. In
particular $N=N(vol(E_i),vol(M_i),\ap,g_s)$ is a function of $g_s$
and the entropy counting considerations go through irrespective
of how $g_s$ is chosen. All what changes when we vary $g_s$ (and
keeping $\ap$ plus the volume of the dual branes fixed) is the
size of the cells and thus their number. The combined cell volume
for one of the dual pairs is given by
\begin{equation}
\text{cell volume}=(\tau_E\tau_M)^{-1}\propto\ap^4g_s^2
\end{equation}
for the string-theory cases and proportional to $l_{Pl}^9$ for the
M-theory case where $E,M=M2,M5$. Therefore in the weakly coupled
string-theory limit, where $g_s\rightarrow 0$, the cell volume
becomes infinitesimally small while the number of cells approaches
infinity $N\rightarrow\infty$ when keeping the brane volume fixed.
The extreme strongly coupled limit $g_s\rightarrow\infty$ on the
contrary would exhibit huge cells covering big portions of the
brane worldvolume.
%(see fig.\ref{cellsize})
%\begin{figure}[t]
%\begin{center}
%\begin{picture}(200,120)(0,0)
%
%\Line(5,15)(5,85)
%\Line(25,25)(25,95)
%\Line(45,35)(45,105)
%
%\Line(85,50)(125,90)
%\Text(220,109)[]{$E$}
%\end{picture}
%\caption{Depicted is the structure of one of the dual pairs $(E,M)$.
%The individual lattice sizes $l_{E,M}$ may (for $Dp$ and $NS5$) or
%may not (in the $F1$ case) depend on $g_s$. However the final cell
%size $v_E v_M=(\tau_E\tau_M)^{-1}$ does.}
%\label{cellsize}
%\end{center}
%\end{figure}

\section{Implications for the Schwarzschild Black Hole}
Let us finally come to some consequences for the Schwarzschild
black hole. We had obtained that ${\cal S}_{BH}$ should equal an
integer $N$. By using the Schwarzschild radius to transform the
horizon area into a mass squared, one sees that for the D=4
Schwarzschild black hole this leads directly to a discrete
mass-spectrum
\begin{equation}
M_{BH}(N)={\cal C}\sqrt{N} \; ,
\quad {\cal C}=\frac{1}{\sqrt{4\pi G_4}}
\label{BMS}
\end{equation}
of Bekenstein-type. This relation has also been found in many
different ways (see e.g.~the references in \cite{K13}). However,
when written in terms of the Schwarzschild radius $r_S$ itself
\begin{equation}
r_S = \frac{l_{Pl}}{\sqrt{\pi}}\sqrt{N} \; ,
\end{equation}
it suggests an effective D=4 picture of the black hole as a random
Brownian Walk with step-width $l_{Pl}/\sqrt{\pi}$. Such an
effective picture in terms of random walks is hard to understand
in many alternative approaches which derive (\ref{BMS}) but
appears to be compatible with the higher-dimensional chain-state
description which is likewise a random walk however with variable
step-width. This and the exact matching of the chain entropy with
the black hole's BH-entropy suggests to identify the D=4
Schwarzschild black hole at a microscopic level with one of the
chain-states and consequently to identify the black hole's mass
with the chain's energy \cite{K13}
\begin{equation}
E_c(N) = M_{BH}(N) \; .
\end{equation}
The importance of this identification lies in the fact that it
allows us to derive thermodynamical quantities within the
microcanonical chain ensemble with given energy $E_c$. Without
this identification one would have to know the complete dynamical
chain-theory in order to derive such quantities.

One obtains a chain-temperature $T_c$
\begin{equation}
\frac{1}{T_c}=\frac{\partial{\cal S}_c(N)}{\partial E_c(N)}
=\frac{1}{T_H}-\frac{1}{M_{BH}} +{\cal
O}\big(\frac{M_{Pl}^2}{M_{BH}^3}\big)
\end{equation}
where
\begin{equation}
\frac{1}{T_H}=8\pi G_4M_{BH} \simeq \frac{M_{BH}}{M_{Pl}^2}
\end{equation}
is the inverse Hawking temperature of the black hole. Hence the
leading-order chain temperature equals the Hawking temperature
but in addition there is a non-trivial correction suppressed by a
factor $M_{Pl}^2/M_{BH}^2$. As expected this correction becomes
important for small black holes whose Schwarzschild radius
approaches the Planck-length. Similarly one can derive the
chain's specific heat as
\begin{equation}
C_c=\frac{\partial E_c(N)}{\partial T_c(N)} = -8\pi G_4 M_{BH}^2
+ 3 +{\cal O}\big(\frac{M_{Pl}^2}{M_{BH}^2}\big)
\end{equation}
where
\begin{equation}
C_{BH}=-8\pi G_4M_{BH}^2
\end{equation}
is the specific heat derived from conventional black hole
thermodynamics. Thus, once more the leading order result
reproduces the black hole thermodynamics result while in addition
there is a correction suppressed by a factor $M_{Pl}^2/M_{BH}^2$.

\bigskip
\noindent {\bf Acknowledgements}\\[2ex]
The author thanks the organizers of the SUSY'02 conference at
DESY, Hamburg and the organizers of the 2002 RTN ``Quantum
Structure of Spacetime'' workshop in Leuven for the opportunity to
present this work. A.K.~is supported by the National Science
Foundation under Grant Number PHY-0099544.

\newcommand{\zpc}[3]{{\sl Z.Phys.} {\bf C\,#1} (#2) #3}
\newcommand{\npb}[3]{{\sl Nucl.Phys.} {\bf B\,#1} (#2) #3}
\newcommand{\npbps}[3]{{\sl Nucl.Phys.B(Proc.Suppl.)} {\bf #1} (#2) #3}
\newcommand{\plb}[3]{{\sl Phys.Lett.} {\bf B\,#1} (#2) #3}
\newcommand{\prd}[3]{{\sl Phys.Rev.} {\bf D\,#1} (#2) #3}
\newcommand{\prb}[3]{{\sl Phys.Rev.} {\bf B\,#1} (#2) #3}
\newcommand{\pr}[3]{{\sl Phys.Rev.} {\bf #1} (#2) #3}
\newcommand{\prl}[3]{{\sl Phys.Rev.Lett.} {\bf #1} (#2) #3}
\newcommand{\prsla}[3]{{\sl Proc.Roy.Soc.Lond.} {\bf A\,#1} (#2) #3}
\newcommand{\jhep}[3]{{\sl JHEP} {\bf #1} (#2) #3}
\newcommand{\cqg}[3]{{\sl Class.Quant.Grav.} {\bf #1} (#2) #3}
\newcommand{\prep}[3]{{\sl Phys.Rep.} {\bf #1} (#2) #3}
\newcommand{\fp}[3]{{\sl Fortschr.Phys.} {\bf #1} (#2) #3}
\newcommand{\nc}[3]{{\sl Nuovo Cimento} {\bf #1} (#2) #3}
\newcommand{\nca}[3]{{\sl Nuovo Cimento} {\bf A\,#1} (#2) #3}
\newcommand{\lnc}[3]{{\sl Lett.~Nuovo Cimento} {\bf #1} (#2) #3}
\newcommand{\ijmpa}[3]{{\sl Int.J.Mod.Phys.} {\bf A\,#1} (#2) #3}
\newcommand{\rmp}[3]{{\sl Rev. Mod. Phys.} {\bf #1} (#2) #3}
\newcommand{\ptp}[3]{{\sl Prog.Theor.Phys.} {\bf #1} (#2) #3}
\newcommand{\sjnp}[3]{{\sl Sov.J.Nucl.Phys.} {\bf #1} (#2) #3}
\newcommand{\sjpn}[3]{{\sl Sov.J.Particles\& Nuclei} {\bf #1} (#2) #3}
\newcommand{\splir}[3]{{\sl Sov.Phys.Leb.Inst.Rep.} {\bf #1} (#2) #3}
\newcommand{\tmf}[3]{{\sl Teor.Mat.Fiz.} {\bf #1} (#2) #3}
\newcommand{\jcp}[3]{{\sl J.Comp.Phys.} {\bf #1} (#2) #3}
\newcommand{\cpc}[3]{{\sl Comp.Phys.Commun.} {\bf #1} (#2) #3}
\newcommand{\mpla}[3]{{\sl Mod.Phys.Lett.} {\bf A\,#1} (#2) #3}
\newcommand{\cmp}[3]{{\sl Comm.Math.Phys.} {\bf #1} (#2) #3}
\newcommand{\jmp}[3]{{\sl J.Math.Phys.} {\bf #1} (#2) #3}
\newcommand{\pa}[3]{{\sl Physica} {\bf A\,#1} (#2) #3}
\newcommand{\nim}[3]{{\sl Nucl.Instr.Meth.} {\bf #1} (#2) #3}
\newcommand{\el}[3]{{\sl Europhysics Letters} {\bf #1} (#2) #3}
\newcommand{\aop}[3]{{\sl Ann.~of Phys.} {\bf #1} (#2) #3}
\newcommand{\jetp}[3]{{\sl JETP} {\bf #1} (#2) #3}
\newcommand{\jetpl}[3]{{\sl JETP Lett.} {\bf #1} (#2) #3}
\newcommand{\arnps}[3]{{\sl Ann.Rev.Nucl.Part.Sci} {\bf #1} (#2) #3}
\newcommand{\acpp}[3]{{\sl Acta Physica Polonica} {\bf #1} (#2) #3}
\newcommand{\sci}[3]{{\sl Science} {\bf #1} (#2) #3}
\newcommand{\nat}[3]{{\sl Nature} {\bf #1} (#2) #3}
\newcommand{\pram}[3]{{\sl Pramana} {\bf #1} (#2) #3}
\newcommand{\hepph}[1]{{\sl hep--ph/}{#1}}
\newcommand{\desy}[1]{{\sl DESY-Report~}{#1}}
\bibliographystyle{plain}

\end{document}